\documentclass[]{aastex631}

\usepackage{comment}
\usepackage{cleveref}

\usepackage{graphicx}
\usepackage[caption=false]{subfig}	

\shorttitle{Absorption features in sub-TeV gamma-ray spectra of BL~Lac objects}
\shortauthors{Foffano et al.}
\graphicspath{{./}{figures/}}

\begin{document}

\title{Absorption features in sub-TeV gamma-ray spectra of BL~Lac objects}

\author[0000-0002-0709-9707]{L. Foffano}
\affiliation{INAF - IAPS, via del Fosso del Cavaliere 100, I-00133 Roma (Italy)}

\author{V. Vittorini}
\affiliation{INAF - IAPS, via del Fosso del Cavaliere 100, I-00133 Roma (Italy)}

\author{M. Tavani}
\affiliation{Astronomia, Accademia Nazionale dei Lincei, via della Lungara 10, I-00165 Roma, Italy}
\affiliation{INAF - IAPS, via del Fosso del Cavaliere 100, I-00133 Roma (Italy)}
\affiliation{Università ``Tor Vergata'', Dipartimento di Fisica, via della Ricerca Scientifica 1, I-00133 Roma, Italy}
\affiliation{ Gran Sasso Science Institute, viale Francesco Crispi 7, I-67100 L’Aquila, Italy}

\author{E. Menegoni}
\affiliation{INAF - IAPS, via del Fosso del Cavaliere 100, I-00133 Roma (Italy)}



\begin{abstract}
The production site of gamma rays in blazars is closely related to their interaction with the photon fields surrounding the active galactic nucleus. 
In this paper, we discuss an indirect method that may help to unveil the presence of ambient structures in BL~Lac objects through the analysis of their gamma-ray spectrum.\\
Gamma~rays, passing through structures at different distances from the black hole, interact via $\gamma\gamma$ pair production with the corresponding photon fields and produce absorption features in their spectral energy distribution.
An interaction with a putative broad-line region may reduce the gamma-ray flux only if its production site were very close to the central engine.
However, if jet photons interact with a bath of \mbox{optical-UV} seed photons produced by a narrow-line region extended over the parsec scale, the consequent $\gamma\gamma$ process may cause absorption features detectable at a few hundreds GeV.\\
The detection of such absorption features is facilitated in sources with spectra reaching TeV energies, and specifically HBLs and EHBLs (\emph{extreme blazars}) may represent exceptional probes to investigate this topic. 
We discuss recent observations of an extreme blazar named 2WHSP~J073326.7+515354 (or PGC~2402248), which shows evidence of such an absorption feature in its gamma-ray spectrum and narrow emission lines in the optical spectrum, suggesting the presence of narrow-line regions in its large-scale environment.\\
Finally, we discuss how sub-TeV absorption features in the spectra of BL~Lac objects may affect their broadband modeling, and eventually represent a powerful diagnostic tool to constrain the gamma-ray production site and the jet environment.
\end{abstract}



\section*{Introduction}
\label{sec:intro}

\noindent
Twenty-five years after the first unified models of active galactic nuclei (AGNs), their precise structure, components, and emission mechanisms are still under debate. In the case of \emph{blazars} - AGNs that emit relativistic jets pointing toward the line of sight of the observer - the presence of ambient photon fields is usually deduced from their optical spectra. Such a method is facilitated in those blazars - called Flat Spectrum Radio Quasars (FSRQs) - in which the optical spectrum is rich in emission lines. In a second category of blazars - called BL~Lac objects - these lines are faint or not present at all \citep[e.g.][]{urry1995, giommi2012}, and the thermal disk components are overwhelmed by the non-thermal radiation of the jet.
These two classes of blazars are probably related by a common evolutionary history \citep[][]{Cavaliere:2002}, and each stage of it may be associated with different properties of the AGN environmental structures. In this paradigm, FSRQs are energized by high accretion rates $\dot{M}\gg 10\%\:\dot{M}_{\text{Edd}}$ ($\dot{M}_{\text{Edd}}$ is the Eddington accretion rate) of cold nuclear gas, which produce powerful and radiatively efficient thermal disk components \citep[\emph{radiative mode},][]{Heckman:2014kza}. The evolution of these objects is fast and depends on the amount of cold gas that accretes onto the nuclear region.  Conversely, BL~Lac objects may represent a late stage of this evolution, with a long-lived phase of accretion fueled only by hot intergalactic gas. The low accretion rates $\dot{M} \lesssim 1\%\;\dot{M}_{\text{Edd}}$ produce radiatively inefficient accretion disks. The power is extracted directly from the rotating black hole \citep{BlandfordZnajek1977} and provided to the particles accelerated in the jet (\emph{jet mode}).

The characterization of structures surrounding BL~Lac jets is also correlated with the production site of the most powerful radiation emitted by blazars, i.e. the gamma rays. The location of their emission has been systematically studied in  FSRQs \citep{costamante-blr}, but a discussion on BL~Lac objects - poorer of optical information  - is still missing.
Our study of the production site of gamma rays in BL~Lacs is performed by investigating their possible interactions with ambient photon fields located along the trajectory of the blazar jet at different distances from the central black hole. Depending on the location of such ambient photon fields, 
$\gamma\gamma$ interactions may take place and cause absorption of jet photons.
In the case of AGN photon fields - that typically have energies ranging from infrared to optical and UV - such $\gamma\gamma$ interactions would produce the strongest effects in gamma rays.
In this regard, BL~Lac objects - and specifically those emitting at the highest energies -  produce large fluxes of gamma rays extending up to the very-high energies (VHE, energies above 100 GeV), with a gamma-ray spectrum not strongly contaminated by other processes or radiative contributions. For this reason, their spectra are ideal probes to verify the presence of external photon fields interacting with the AGN jet.

BL~Lac objects represent an interesting type of sources with scarcely contaminated non-thermal emission of the jet that produces a spectral energy distribution (SED) mainly composed of two humps. The first hump is commonly interpreted as synchrotron radiation emitted by relativistic electrons moving in the jet magnetic field.
The second component peaks at VHE gamma rays and its origin is more debated. In the so-called Synchrotron Self-Compton model \citep[SSC, e.g.][]{Maraschi:SSC, Tavecchio:SSC}, it is mainly produced by inverse Compton scattering of low-energy target photons \citep{IC} coming from the same electrons producing the synchrotron hump.

BL~Lac objects are classified on the basis of their synchrotron peak frequency $\nu_{\text{peak}}^{\text{sync}}$ \citep[according to][]{2010ApJ...716...30A}. In this paper, we will consider the high-peaked BL~Lac objects (HBL, $\nu_{\text{peak}}^{\text{sync}}$ between $10^{15}$ and $10^{17}$ Hz) and the extremely high-peaked objects (EHBLs or also \emph{extreme blazars}), which have been defined when $\nu_{\text{peak}}^{\text{sync}} > 10^{17}$~Hz, extending in \citet{foffano-2019} the definition provided in \citet{2010ApJ...716...30A}.
Specifically, the latter is an emerging class of very rare BL~Lac objects \citep[][]{costamante_2001}, usually pretty difficult to be studied due to their lower luminosities  in the hard X-ray band (where the first hump peaks) 
and at gamma rays (where the second hump lies).
However, as a consequence of an increasing number of broadband observations of these sources,  this population has increased in the last years, revealing interesting behaviours and spectral differences appearing in their phenomenology \citep[][]{foffano-2019}.

In unified models of AGNs, BL~Lac objects have been associated with some specific types of radio galaxies. This association may be useful to increase the statistics and provide a richer context for the investigation. Radio galaxies are radio-loud AGNs with jets pointed at large angles with respect to the observer's line-of-sight. They were classified for the first time by \citet{Fanaroff_Riley_1974} on the basis of radio morphological criteria, which provided two different classes: FR-I objects with high surface brightness near the core and along the jets, and FR-II objects with higher surface brightness on the lobes located far from the central region. 
Specifically, BL~Lac objects are associated with FR-I radio galaxies \citep[e.g.][]{urry1995}.
A more recent classification of radio galaxies is based on their optical spectra, and considers the emission lines produced in their narrow-line regions \citep[][]{Laing1994}: high-excitation radio galaxies when they are strong (HERGs, equivalent width~[O III] $>3$~\AA), and low-excitation radio galaxies when the lines are weak (LERGs).
In this more modern classification of radio galaxies, BL~Lac objects are associated to LERGs, which come both as FR-I and FR-II objects \citep[e.g.][]{Macconi:2020joo}.\\

\noindent
In Section~\ref{sec:gammagammainteraction} of this paper we will describe the $\gamma\gamma$ interaction and its absorption effects on a given photon flux. In Section~\ref{sec:absorptionfeatures}, we will apply such an interaction to photons coming from the most important ambient structures surrounding the jet, i.e. a hypothetical broad-line region (BLR) and a more external photon field, like a narrow-line region (NLR). We will discuss the spectral consequences weather the gamma-ray production site were behind the BLR (Section~\ref{sec:blr_in_bllac_objects}) or between the BLR and the NLR (Section~\ref{sec:nlr_in_bllac_objects}). In Section~\ref{sec:pgc_data}, we will show a possible evidence of this process in the gamma-ray spectrum of a recently TeV-detected BL~Lac object. Finally, in Section~\ref{sec:discussion} we will provide a detailed discussion and prospects of this method.\\



\section{The gamma-gamma absorption process}
\label{sec:gammagammainteraction}

\noindent
The $\gamma\gamma$ interaction takes place when two photons collide and produce an electron-positron pair. In this paper, we will consider highly energetic jet photons with energy $\epsilon_{\gamma\text{-ray}}$ that collide with a bath of seed photons with lower energy $\epsilon_{\text{seed}}$.
The $\gamma\gamma$ absorption process is possible only when - in the lab frame -  the energies of the incoming photons satisfy \citep[for more details see][]{aharonian2004book}:
\begin{equation}
\epsilon_{\text{seed}} \cdot \epsilon_{\gamma\text{-ray}} \ge \frac{2\, \Big(m_e c^2\Big)^2}{1- \cos\theta} \sim \frac{5.2\cdot10^{11}\, {\rm eV}^2}{1 - \cos \theta} \; ,
\label{eq:gammagammathreshold}
\end{equation}
where $\theta$ is the angle between the trajectories of the two photons,
$m_e$ is the electron mass and $c$ is the speed light.

\begin{figure}
\centering
\includegraphics[width=0.75\textwidth]{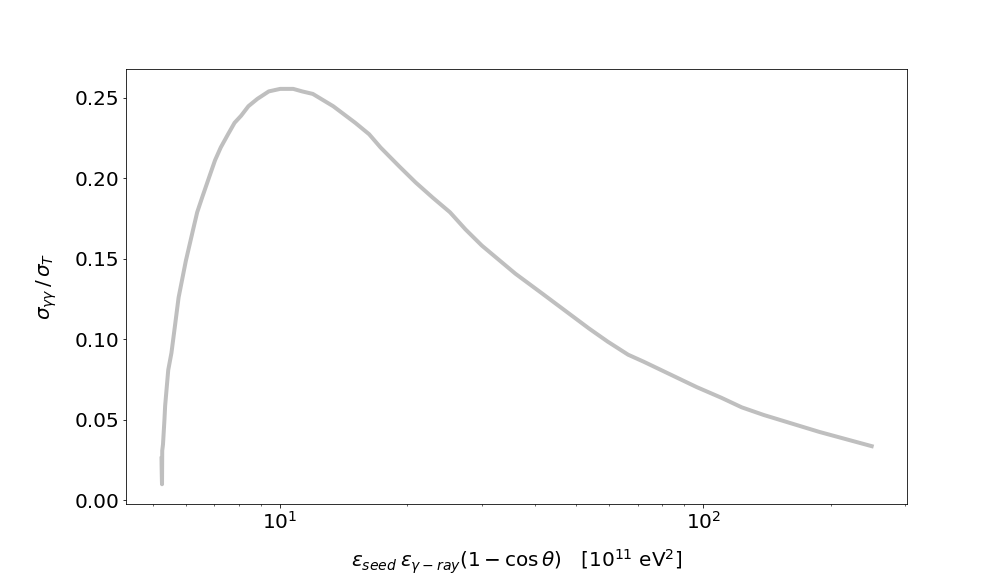}
\caption{Cross section of the $\gamma\gamma$ interaction producing electron-positron pairs in units of the Thomson cross section $\sigma_T$ and as a function of the product of the interacting photon energies (see \Cref{eq:gammagammathreshold}).}
\label{fig:gammagammacrosssection}
\end{figure}

\noindent
The $\gamma\gamma$ cross section in the mono-energetic isotropic photon field is represented in \Cref{fig:gammagammacrosssection} as a function of the energy of the incoming photons. The threshold for the interaction is the lowest in the case of head-on collisions with $\theta = \pi$.
The cross section reaches the maximum value at twice the threshold energy very quickly, and afterwords its value decreases as the energy of the incoming gamma-ray photons increases. Considering for simplicity only narrow mono-energetic distributions of seed photons, we expect that the effect on the spectrum would be a sharp reduction of the flux at the corresponding threshold energy of the reaction, and then a gradual reduction of this absorption effect along the following order of magnitude in energy.

When the blazar jet encounters a bath of seed photons and the $\gamma\gamma$ process takes place, the interacting photons produce electron-positron pairs that reduce the observed flux $I_{\text{out}}$ with respect to the incoming photons flux $I_{\text{in}}$.
This process can be described by:
\begin{equation}
I_{\text{out}}  = I_{\text{in}} \: e^{-\tau_{\gamma\gamma}} \; ,
\label{eq:absorption-flux}
\end{equation}
where $\tau_{\gamma\gamma}$ is the absorption factor.
As a general rule, the absorption factor can be expressed as a function of the cross section $\sigma_{\gamma\gamma}$ of the $\gamma\gamma$ interaction, of the size of the interacting region $R$, and of its photon density $n_{\text{seed}}$. If we assume a mono-energetic, isotropic, and uniform seed photon field, it becomes:
\begin{equation}
\tau_{\gamma\gamma} = \; n_{\text{seed}} \cdot \sigma_{\gamma\gamma} \cdot R \; \simeq \;0.68 \cdot  n_{\text{seed,4}} \cdot (R/100\text{ pc})\; ,
\label{eq:tau-definition}
\end{equation}
where we adopt the notation $n_{\text{seed}} = 10^4 \: n_{\text{seed,4}}$.
It is worth to note that, assuming a constant cross section over the energy of the interacting photon fields, the absorption factor $\tau_{\gamma\gamma}$ is mainly dependent on the photon column density $K_{\text{seed}} = n_{{\text{seed}}} \cdot R$ (with notation expressed in $\frac{pc}{cm^3}$ to explicit the typical values of size and photon density).

This process can be obviously complicated when the seed photons are distributed over a wide range of energies, enlarging the cross section of the interaction over a wider energy range. Additionally, we are simplifying the discussion assuming a uniform density of the seed photon field, but this could be changing over long distances and produce a non uniform absorption process.\\


\section{Absorption features in gamma-ray spectra of BL~Lac objects}
\label{sec:absorptionfeatures}

\noindent
The most important structures of material surrounding the AGN and the blazar jet are the broad-line region (BLR) and the narrow-line regions (NLRs). 
Depending on the location of the gamma-ray emission with respect to these structures and on the source that illuminates them, the respective re-emitted soft photons can interact and produce different spectral features in the observed gamma-ray spectrum. \\

\subsection{Interaction with the BLR}
\label{sec:blr_in_bllac_objects}

\noindent
Recent studies concerning the BLR show that it is probably contained in a region  close to the central black hole up to about $R_{\text{BLR}} \simeq 0.1$ parsec from it \citep[][]{Peterson2006}. Most of the modern BLR models represent it as a set of high particle density ($\gtrsim 10^9$~cm$^{-3}$) and hot \citep[$\sim 10^5$ K,][]{tavecchio2008-blr} clouds, which are distributed around the central source with a covering factor $f\gtrsim 10\%$ (i.e., fraction of the sky covered by BLR clouds). These clouds are illuminated by the continuum of the accretion disk or partially by the jet itself \citep[see][]{ghisellinimadau}, and have typical reflectivity $a\sim$ 30\%: the total reprocessing factor of the radiation investing the clouds results $\alpha = a \cdot f \sim $ few $10^{-2}$.
They are rotating rapidly at several thousands km/s and then emitting lines broadened by the Doppler effect in the optical spectrum \citep[][]{osterbrock}.

Assuming that these clouds are isotropically distributed around the accretion disk producing the ionizing luminosity~$L_d$, and that they re-emit through emission lines \citep[the most important being the Ly$_\alpha$, see][]{poutatenstern2010, Stern:2014uqa} from distances $R_{\text{BLR}}$, then the photon density at $R\le R_{\text{BLR}}$ is
\begin{equation}
    n_{\text{BLR}}\approx \frac{\alpha\,L_d}{4\pi\,c\,R^2_{\text{BLR}}\epsilon_{\text{seed}}}   \; ,
    \label{eq:blr_photon_density}
\end{equation}  
where $\epsilon_{\text{seed}} \simeq 13\,$eV is the typical energy of BLR photons.
From reverberation maps \citep[see][]{BlandfordMcKee1982, Peterson1993_reverberation_mapping,Bentz2006,kaspi2007}, a correlation between line luminosities and $R_{\text{BLR}}$ \citep[e.g.][]{Koratkar1991} -  with the assumption $\alpha\simeq$ const - indicates a simple relation \citep{blazarsequence08}
\begin{equation}
 R_{\text{BLR}}\simeq\,10^{17} L_{d,45}^{0.5}\:\text{cm}  \; .
 \label{eq:blr_radius_vs_luminosity}
\end{equation}
In this way, the photon density into the BLR results $n_{\text{BLR}}\approx 10^8\,\alpha_{-2}$ cm$^{-3}$.

If gamma rays are produced behind the BLR, when passing through it they would suffer $\gamma\gamma$ interaction with its dense photon fields. From  \Cref{eq:tau-definition}, the resulting absorption factor $\tau\simeq 7\;\alpha_{-2}\;(R_{\text{BLR}}/0.1\text{ pc})$ over a BLR radius of $\sim0.1$~pc would strongly affect the observed spectrum of FSRQs already above some tens of~GeV.
This process has been tested by \citet{costamante-blr} in the case of FSRQs detected with \emph{Fermi}-LAT \citep{Fermitelescope}, where the authors find that most of them do not show any strong spectral absorption in the energy range interested by such a process. This means that either the gamma-ray emission is located outside the BLR region, or that the BLR is much different from our expectations, specifically about its extension and location \citep[about $100\times$ larger than given by reverberation mapping, e.g.][]{Peterson2006}, or that the reprocessed fraction $\alpha$ is sensibly less than 5\%.

Differently from the discussion of \citet{costamante-blr} for FSRQs, in BL~Lac objects the accretion is low: a putative disk would have luminosity $\lesssim 10^{43}$~erg s$^{-1}$ and from \Cref{eq:blr_radius_vs_luminosity} 
the respective BLR would lie at distances $\lesssim10^{16}$~cm from the central engine, so 
$\tau_{\gamma\gamma}\simeq0.2\;\alpha_{-2}\;R_{\text{BLR},16}$. In this case, in BL~Lac objects we could not exclude some absorption of gamma~rays in a very inner BLR ($R_{{\text{BLR}}}\lesssim 10^{16}$~cm) when $\alpha\gtrsim 3$\%. Only the gamma-ray component emitted in this very inner region would be absorbed, whereas those emitted farther would not. So, no detectable $\gamma\gamma$ absorption would emerge if high-energy components are produced also beyond $10^{16}$~cm.

In this paper, we will then explore when the gamma-ray emission is produced outside the region possibly occupied by the BLR, and in that case such an energetic radiation would be free to escape and to continue its travel through the blazar jet, eventually interacting with other photon fields. \\



\subsection{Interaction with narrow-line regions}
\label{sec:nlr_in_bllac_objects}

\noindent
We will now discuss the possible $\gamma\gamma$ interaction of the blazar jet with photon fields at larger distances of several parsec from the central engine, considering the gamma-ray emission being produced also outside the region of influence of the BLR. 

Among the structures composing the AGN, the NLR is the most extended one. Even though its main properties are not well constrained, thanks to optical studies on radio galaxies the NLR has been spatially resolved in many objects. 
Its morphology is probably axisymmetrical (rather than spherical), represented with a conical outflow of material 
collimated by the torus \citep[e.g.][]{Peterson2006, Schweitzer:2008ev}. Its size $R$ - whose limits may depend on the specific AGN - ranges between 1 and a few 10$^2$~parsec from the central engine \citep[see][]{osterbrock}. In some sources, the evidence of further extensions of this region above such scales needed the definition of an \emph{extended NLR} (ENLR), lying from $\sim 100$ pc to 10 kpc \citep[e.g.][and references therein]{Congiu2017}.

The NLR is mostly composed by clouds of dust and gas with electron densities ranging between $10^2$ and $10^4$~cm$^{-3}$.
The electron densities are lower with respect to the BLR: for this reason, forbidden lines are not collisionally suppressed and do appear in the optical spectrum of these sources. The relations between these emission lines provide estimates of the electron temperature with a typical value at $T\simeq 1.6\cdot10^4$~K \citep[][]{Koski1978}.
These clouds are rotating at a few hundreds km/s, with average covering factor of $f\sim 16\%$ \citep[e.g.][]{Schweitzer:2008ev}, and produce narrow lines in the optical spectrum.
The NLR is characterized by a wide range of ionization levels and emitted radiation ranging from optical to UV energies. Among them, high-ionization lines of [O III]$\lambda5007$ are usually considered a good indicator of the NLR and tracers of outflows over 10~pc scales \citep[][]{Macconi:2020joo,Singha2021}.

\begin{figure}
\centering
\includegraphics[width=0.7\textwidth]{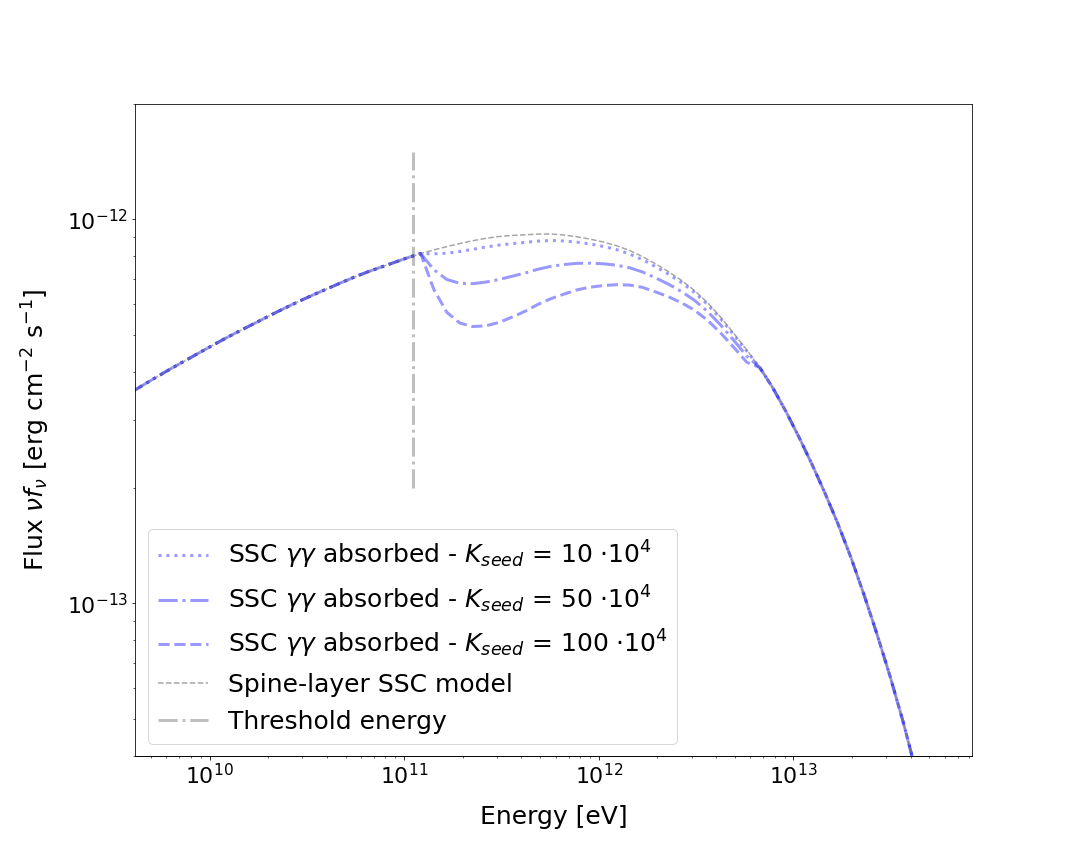}
\caption{Simulated observed gamma-ray spectrum (already virtually EBL de-absorbed) of a typical EHBL affected by $\gamma\gamma$ absorption resulting from interaction of photons of the blazar jet with seed photons of energy $\epsilon_{\text{seed}}\simeq4$~eV. We show several hypothetical curves related to different photon column densities $K_{\text{seed}}$ of the seed photon fields.}
\label{fig:generic_gammaray_spectrum_with_NLR_feature}
\end{figure}

Let us assume the presence of a NLR (motivated in Section~\ref{sec:discussion}) and compute its spectral effect.
Following \Cref{eq:gammagammathreshold}, the absorption due to $\gamma\gamma$ interaction with a bath of such optical-UV photons produces an absorption feature in the gamma-ray spectrum of blazars.
An example is reported in \Cref{fig:generic_gammaray_spectrum_with_NLR_feature}, where we show how the observed spectrum of a hypothetical EHBL would be modified by the absorption feature due to three different values of the photon column density $K_{\text{seed}}$. The spectrum is already de-absorbed by the interaction with the extragalactic background light \citep[EBL, e.g.][]{Hauser:2001, Franceschini08, Franceschini17}. The higher is $K_{\text{seed}}$, the deeper is the absorption feature in the gamma-ray spectrum. 
On the other hand, larger energies of the seed photons correspond to a lower threshold of the $\gamma\gamma$ interaction and a shift of the absorption feature to lower energies (and \emph{vice versa}). \\



\section{Absorption features in the gamma-ray spectrum of the BL~Lac PGC~2402248}
\label{sec:pgc_data}
\noindent
An interesting publication by the MAGIC Collaboration \citep[][hereafter PaperI]{MAGIC_paper_pgc} has announced the detection of a new TeV gamma-ray source named PGC~2402248 (also named 2WHSP~J073326.7+515354).
The campaign of simultaneous multi-wavelength (MWL) observations provided by PaperI  confirmed that this source is a new EHBL with synchrotron peak located at $\nu_{\text{peak}}^{\text{sync}} \simeq {10^{17.8 \pm 0.3}}$ Hz and gamma-ray spectrum extending up to 5~TeV. 
The redshift of $z = 0.065$ was reported for the first time by \citet{pgc_optical_spectrum} - hereafter PaperII - with a specific optical campaign.

\begin{figure}
\centering
\includegraphics[width=0.7\textwidth]{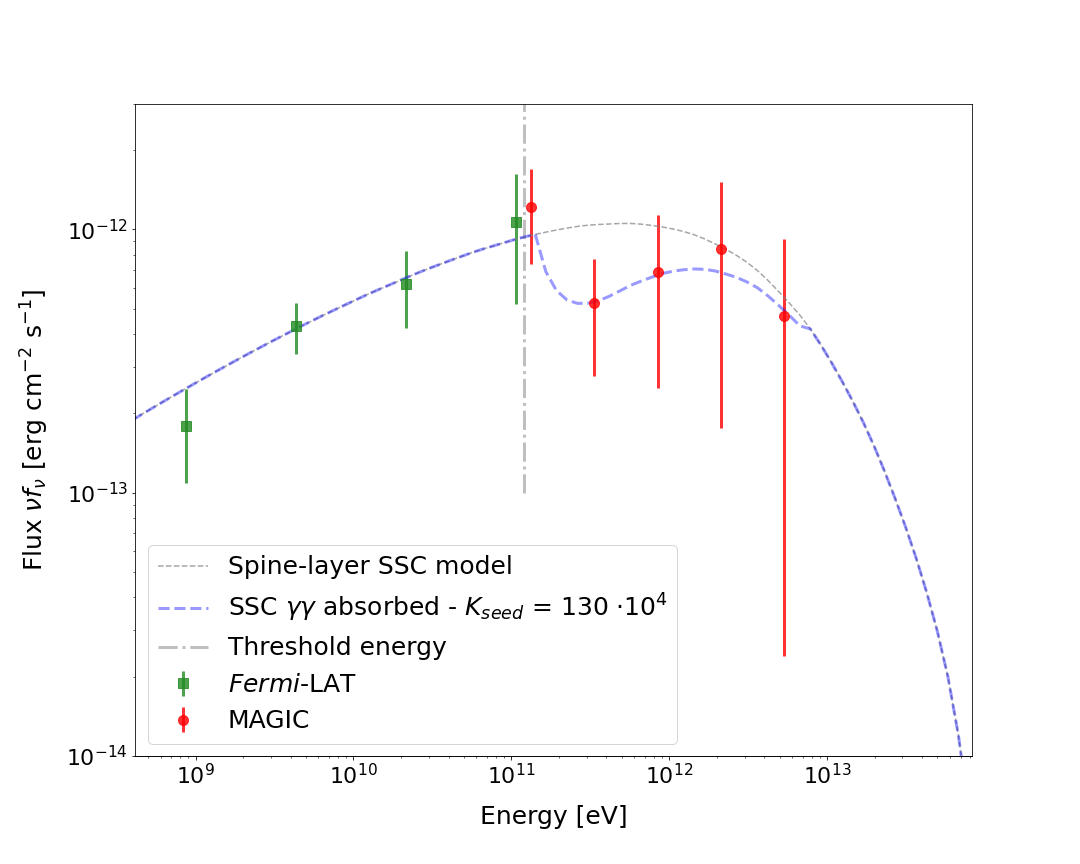}
\caption{Zoom on the gamma-ray data (\emph{Fermi}-LAT data in green squares and MAGIC data in red circles) of the MWL SED of PGC~2402248 \citep[adapted from][]{MAGIC_paper_pgc}. We show also a SSC spine-layer model (grey dotted curve) and the additional $\gamma\gamma$~absorption resulting from the interaction with seed photons of energy $\epsilon_{\text{seed}}\simeq 4$ eV (blue dashed line), happening above the threshold energy (vertical line).}
\label{fig:pgc_gammaray_data}
\end{figure}

In \Cref{fig:pgc_gammaray_data} we show a focus of the MWL data of PaperI around the second hump of the SED, already \mbox{de-absorbed} by the interaction with the EBL using the model proposed by \citet{Dominguez:2010bv}.
The gamma-ray SED composed by high-energy (\emph{Fermi}-LAT, in green squares) and VHE data (MAGIC, in red circles) shows a continuously increasing flux up to about 100~GeV. At that point, \emph{Fermi}-LAT and MAGIC data are well compatible at about 100~GeV with $\text{Flux}_{100\text{GeV}} = (1.06 \pm 0.54) \cdot 10^{-12}$ erg/cm$^{2}$/s, but at slightly higher energies of $\sim 250$~GeV the flux decreases to $\text{Flux}_{250\text{~GeV}} = (0.52 \pm 0.03) \cdot 10^{-12}$ erg/cm$^{2}$/s. Then, the flux grows up again gradually to rejoin the expected SSC emission at about a few TeV.

\begin{figure}
\centering
\includegraphics[width=0.85\columnwidth]{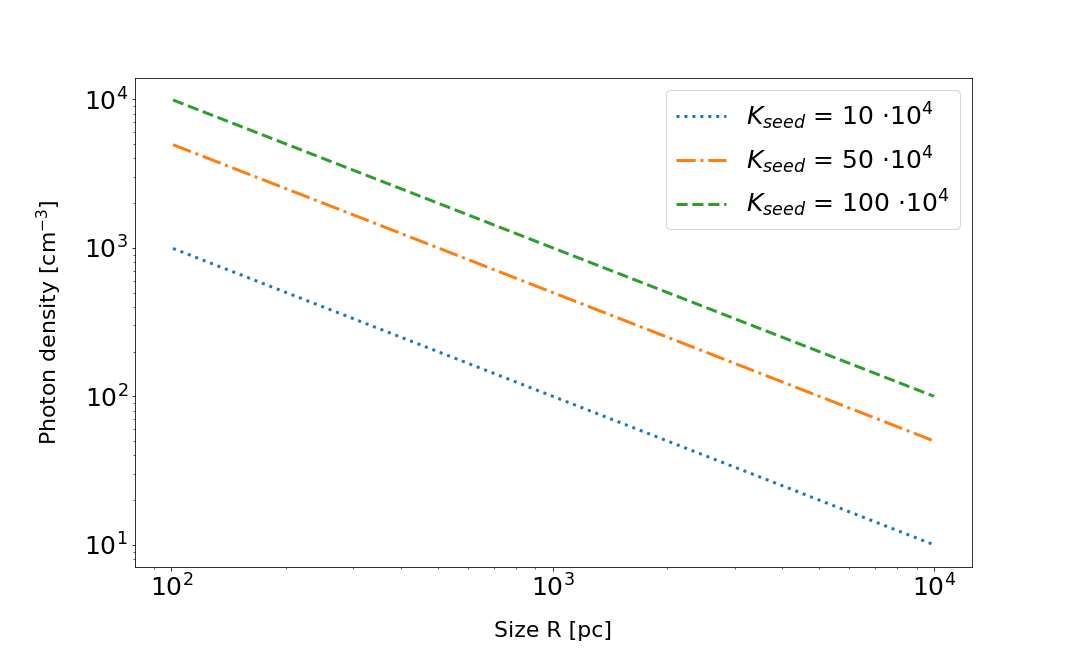}
\caption{Different combinations between the extension of the interacting region and its photon density, given the measured photon column density $K_{\text{seed}}$.
}
\label{fig:nlr_product_PGC}
\end{figure}

It is important to note that, due to the different integration time of the observations performed with \emph{Fermi}-LAT and MAGIC in PaperI, such a spectral feature may be actually due to different states of the sources.
However, considering the lack of flux variability found in the available dataset of PaperI,  the usually very low flux variability of EHBLs, and the compatibility of the first MAGIC point with the last \emph{Fermi}-LAT point in the gamma-ray spectrum (see \Cref{fig:pgc_gammaray_data}), we assume that the flux variability of the source is negligible, even if it is an important point to consider in the discussion.\\

\vspace{20pt}

\subsection{Possible effect of a narrow-line region?}
\label{sec:pgc_nlr}

\noindent
The absorption feature of the gamma-ray spectrum of PGC~2402248 may be related to the interaction of the jet with photons emitted by the NLR.
Besides the prominent non-thermal continuum of the jet, the well-sampled optical spectrum of PaperII shows the radiation emitted by the host galaxy, which provides information on the source redshift and on the properties of the population of stars composing the galaxy.
These features are typical for giant elliptical galaxies hosting blazars. Specifically, the galaxy is characterized by old stellar populations showing high metallicity and no evidence of ongoing star formation, with the latest star formation episodes happening at least 6~Gyr ago.

After removing the emission of the host galaxy, the optical spectrum shows several surviving emission lines. The diagnostic diagrams of \citet{Kewley2006} applied to these lines support the classification of this source as a LINER (galaxy with low-ionization nuclear emission-line region) of the AGN type. 
Additionally, following also the discussion of Section~\ref{sec:nlr_in_bllac_objects}, the ratios between the fluxes of the available emission lines are compatible with the definition of LERG as reported in \citet{Buttiglione:2009hp}.
All this information - considering that ongoing star formation has been excluded - provides support to the presence of a NLR producing the narrow emission lines.

Let us now compute the main parameters describing the spectral feature of \Cref{fig:pgc_gammaray_data}, assuming that it is due to $\gamma\gamma$ absorption applied to an underlying spectral SSC emission of a two-zone spine-layer model \citep[][such as in PaperI]{Ghisellini05}.
For continuity of the spectral emission in the gamma-ray spectrum, we can suppose that the input flux before entering the absorption region at 250~GeV was $I_{\text{in}} \equiv \text{Flux}_{100\text{GeV}}$ and that the observed flux is $I_{\text{out}} \equiv \text{Flux}_{250\text{GeV}}$. Then, from \Cref{eq:absorption-flux} and assuming that the absorption feature peaks at $\sim 250$~GeV, we obtain a rough estimate of the maximum absorption factor of $\tau_{\gamma\gamma} = 0.70 \pm 0.36$.
Following \Cref{eq:tau-definition}, the total photon column density that may produce this absorption factor is $K_{\text{seed,max}} = (104\pm37) \cdot 10^4 \; \frac{pc}{cm^3}$.

A better determination of the physical parameters can be obtained by fitting the underlying SSC emission - absorbed by $\gamma\gamma$ absorption -  and releasing the assumption that the absorption feature peaks at $\sim 250$~GeV. This means that we can vary the energy of the photon field (and consequently the energy of the maximum absorption), looking for the values of the seed photon fields energy and column density that minimize the chi-square of the fit. Such an analysis reports an interval of the energy of the seed photon fields $\epsilon_{\text{seed}}$ between about 1 and 4~eV, which includes a bath of seed photons in the optical-UV.
These values are compatible with the strongest emission lines of the spectrum of  well-studied narrow-line radio galaxies \citep[such as \mbox{Cyg-A},][]{osterbrock} and also with the optical spectrum of PaperII.
Then, assuming a narrow distribution of mono-energetic seed photons with energy near 4.0~eV, the value of the photon column density $K_{\text{seed}}$ that minimizes the chi-square is $K_{\text{seed}} \simeq 130 \cdot 10^4 \,
\frac{pc}{cm^3}$ (and similar values are reported also for the other values of $\epsilon_{\text{seed}}$). It is interesting to note that all photon column densities are in agreement with the value $K_{\text{seed,max}}$ obtained before.

In \Cref{fig:nlr_product_PGC} we show - for the three example values of $K_{\text{seed}}$ used in \Cref{fig:generic_gammaray_spectrum_with_NLR_feature} - the corresponding parameter space of the two variables (size and photon density) of the interacting region.
The experimental evaluation of the photon column density $K_{\text{seed}}$ -  compatible with the green line - allows us to limit this parameter space and provide experimental constraints on the properties of the narrow-line regions.\\

\subsection{What is the origin of the required photon density?}
\label{sec:photon_density}

\noindent
An important point concerning the possible presence of a NLR is the origin of the photon column densities required to produce the absorption factor found in Section~\ref{sec:pgc_nlr}.
Let us consider a resulting photon column density of $K_{\text{seed}} \simeq 100 \cdot 10^4 \; \frac{pc}{cm^3}$, represented by the green dashed line shown in \Cref{fig:nlr_product_PGC}. Typical results on narrow-line regions in literature indicate an average size of the ENLR of 10~kpc and of the NLR around 100~pc, and consequently the required photon densities range between $10^2$ and $10^4$~cm$^{-3}$, respectively.

Similarly to the most common models of BLR radiation, the NLR may be photoionized by an accretion disk.
Considering the formula describing the energy density of the BLR, we can apply the same relation to a distribution of clouds compatible with the NLR:
\[
U_{\text{NLR}} = \frac{a \: f \:L_d}{4\pi c \: R^2} = 2.8 \cdot 10^{-12} \: L_{\text{d,43}}\Bigg[\frac{a \cdot f}{10^{-2}} \Bigg] \Bigg[\frac{R}{100\text{ pc}}\Bigg]^{-2} \; \text{erg cm}^{-3} \; ,  
\]
where $L_d$ is the disk luminosity, $a$ the reflectivity $a\sim$ 50\%, and $f$ the covering factor. The formula is shown also with reference values for a disk luminosity that from real data must be lower than $L_d \lesssim 10^{43}$ erg/s and an indicative cloud distance of $R \simeq 100$~pc. Considering photons of energy $\epsilon_{\text{seed}} \simeq 5$~eV, the corresponding average photons density $n_{\text{seed}}$ becomes
\[
n_{\text{seed}} =\frac{U_{\text{NLR}}}{\epsilon_{\text{seed}}} =  \frac{1.7}{\epsilon_{\text{seed}}} \: L_{\text{d,43}}\Bigg[\frac{a \cdot f}{10^{-2}} \Bigg] \Bigg[ \frac{R}{100\text{ pc}} \Bigg]^{-2} \lesssim \; 2\; \text{cm}^{-3} \; .
\]
Unfortunately, this value of the photon density is not compatible with the parameter space provided by the green dashed line shown in \Cref{fig:nlr_product_PGC} and the discussion made before. Specifically, in order to have possible sizes of the NLR (and of the ENLR) compatible with the observations, they should produce a photon density $n_{\text{seed}}  \gg 10^2\; \text{cm}^{-3}$. Considering that this estimation is conservative and that the actual value of the photon density may be much lower, we conclude that the contribution given by an accretion disk is not sufficient to illuminate the NLR of this object and produce the required photon densities.

As an alternative scenario, considering that in BL~Lac objects the power emitted by the accretion disk is supposed to be marginal, another contribution to the illumination of the NLR may come from the radiation emitted by the jet itself \citep[e.g.][]{ghisellinimadau, vittorini2014}. In fact, considering the synchrotron radiation emitted before or while passing through the NLR, it would radiate photons and illuminate its clouds, that in turn may re-emit and increase locally the seed photon density. Such photon density may be enough for the $\gamma\gamma$ process with the gamma~rays of the jet to happen, and produce the observed reduction of the gamma-ray photon flux.

The jet of typical BL Lac objects emits an intrinsic synchrotron luminosity $L'_{\text{s}}\simeq10^{39}$~erg~s$^{-1}$. Assuming that a system of NLR clouds surrounds the jet at an impact parameter $D\sim 10^{16}$~cm and is illuminated by the jet itself, then from \citet{vittorini2017} the local density of photons in the lab frame would be
\[
n_{\text{j,NLR}} \simeq \frac{f \:L'_s \Gamma^2}{2\:c\, D^2\:\epsilon_{\text{seed}}} \approx 2 \cdot 10^{8} \; {L'_{\text{s,39}}} \:\Bigg[\frac{f}{0.1} \Bigg] \Bigg[\frac{D}{10^{16}\text{ cm}}\Bigg]^{-2} \Bigg[\frac{\Gamma}{10}\Bigg]^2 \Bigg[\frac{\epsilon_{\text{seed}}}{5 \text{ eV}}\Bigg]^{-1} \; \text{cm}^{-3} \; .
\]
This scenario, as demonstrated in \citet{vittorini2014}, provides higher photon densities thanks to the boost given by the moving plasmoid and to its lower distance from the reflecting clouds. In this case, considering as effective interaction path the size of the system $D$ and the specific parameters of this source, the resulting absorption factor $\tau_{\gamma\gamma}\simeq \sigma_{\gamma\gamma}\cdot n_{\text{j,NLR}}\cdot D\approx\,0.5$ would be compatible with that required in Section~\ref{sec:pgc_nlr}. 
Moreover, the case in which clouds surround the jet and are illuminated by the latter, would favour $\gamma\gamma$ interaction at fixed angles close to $\frac{\pi}{2}$ (or, in other words, at minimal distance between clouds and jet) where the dilution of the clouds' re-emission is small: this non-isotropic geometry is conducive to a sharp dip in the energy spectrum, as observed in PGC~2402248.
This scenario requires a deeper investigation, and it will be studied in a dedicated publication.\\

\vspace{10pt}

\section{Discussion}
\label{sec:discussion}

\subsection{BLR and NLR}
\noindent
The absence of evidence for an interaction between gamma rays and a BLR in BL~Lac objects is in agreement with their evolutionary history mentioned before \citep[][]{Cavaliere:2002, Heckman:2014kza} and their association with a class of radio galaxies, the LERGs (Section~\ref{sec:blr_in_bllac_objects}). On the observational point of view, we do not see any indications of BLR typical emission in the optical spectrum of BL~Lac objects and in their broadband SED. 
This implies that, either 1) the BLR is not present at all, or that 2) the BLR is present but it is not adequately photoionized by a low-luminosity or absent accretion disk (typically inferred in LERGs) and that the lines are overwhelmed by the continuum, or that 3) it is strongly different from the typical BLR that we know (for instance, characterized by lower particle density or located at larger distances from the black hole).

A more complicated discussion regards the possible presence of narrow-line regions in BL~Lac objects.
In general, the determination of its presence is challenging due to the scarce appearance of emission lines in their optical spectrum, as they are usually overwhelmed by the non-thermal continuum of the jet. 
However, like in the case of PGC~2402248, some sources may present favourable spectral properties that allow us to investigate the possible presence of a NLR.

Also in this case, the comparison of BL~Lac objects with LERGs may be helpful, providing larger statistics and a richer context for the investigation.
The LERGs constitute a large population of radio galaxies with narrow emission lines in their optical spectra (LINERs are also likely part of this population). 
It is important to notice that these emission lines may be produced also by star formation processes, for example by massive clouds of ionized [H~II] regions \citep{Baldwin1981}. However, when detailed optical observations are available for the host galaxies of blazars, the analysis of their stellar populations should indicate their age and the last evidence for star formation episodes.
Specifically, in most of the galaxies hosting BL~Lac objects there is no evidence for substantial recent star formation, and a notable portion of BL~Lac objects may really present narrow-line regions. Such narrow-line regions may be less powerful than usual due to the lower luminosity of the accretion disk and of the jet able to photoionize  their gases.\\

\subsection{Limitations and opportunities of our method}
\noindent
Our indirect method allows us to detect the presence of an external photon field (e.g. given by a NLR) only when its physical parameters are sufficient to produce a substantial $\gamma\gamma$ absorption in the  BL~Lac gamma-ray spectrum.
In fact, looking at \Cref{fig:generic_gammaray_spectrum_with_NLR_feature} and considering the usually large error bars of the gamma-ray data characterizing faint HBLs or EHBLs, a photon field on the jet trajectory could produce a detectable effect only with sufficiently high photon column densities $K_{\text{seed}}$ of the seed photon fields, i.e. with proper values of the photon density and of the size of the interacting region. Otherwise, the spectral feature could be indistinguishable from the fluctuation of the data points or their error bars. 
In the case of unusual regions  - BLR or NLR with properties different from the typical values mentioned before - the physical results reported by this method remain unaffected, as it estimates the total photon column density of the seed photon field.
When the spectral feature is identified with this method (and then it is possible to evaluate the total photon column density $K_{\text{seed}}$), other complementary methods could estimate independently the size of the candidate NLR or its composition.

Then, it is important to underline that this absorption effect can be detected only in sources with favourable spectral properties, otherwise it may be undetectable (if the spectrum does not extend up to several hundreds GeV) or confused with other features (e.g. the spectrum is already very steep at those energies). 
The best targets are represented by sources that 1) are well detected in VHE gamma rays, 2) show hard spectra extending up to several hundreds GeV, 3) show a clean spectral shape (without contamination of other spectral features) in that band,  4) and show a relatively stable flux at those energies.

Interestingly, extreme blazars (besides standard HBLs) may represent a very powerful tool to investigate the presence of NLR in BL~Lac objects.
In fact, they are characterized by hard spectra extending up to the VHE, and their spectral shape  - once de-absorbed by EBL - is usually very regular (like power-law or log-parabola).
Additionally, their flux variability is very low (except for somw \emph{flaring HBL-like} EHBLs, as reported in \citet{foffano-2019}, but also in that case the spectral shape is usually well defined and stable for long times).

The current populations of TeV-detected EHBLs are located only at quite low redshift ($z < 0.2$), due to their low luminosity (barely reaching the sensitivity of the current gamma-ray detectors) and the strong EBL absorption that unavoidably affects the spectra of sources located at larger  distance. 
The low redshift and the synchrotron emission moved at hard X-ray energies allow the host galaxy to be well-identifiable in the SED from the continuum.
For this reason, dedicated observations of their optical emission may be extremely helpful to understand how many of these sources do present optical emission lines, and then when this is correlated also to a gamma-ray absorption in their second hump of the SED.

Additionally, in recent observations of several EHBLs a sensible excess in the optical-UV has been found. This excess emerges when modeling the SEDs by applying both one-zone and two-zone SSC models \citep[see][]{Nustar_EHBLs}.
The hard high-energy gamma-ray spectra peaking at several TeV implies a corresponding hard synchrotron spectrum raising from radio up to hard X-ray energies. However, in several sources (like 1ES~0229+200 and 1ES~0347-121) the emission at UV wavelengths predicted with such models  is lower than the values actually measured \citep[e.g.][]{tavecchio2009}. \citet{Nustar_EHBLs} suggests that the UV excess may be due to a different synchrotron component (not reprocessed with the SSC), or to a less feasible burst of star formation activity. As described in PaperII, the latter hypothesis can be excluded for PGC~2402248, and similar analyses on other sources may provide the same conclusions. This implies that this UV excess may be correlated with the photon fields produced by a possible NLR. Alternatively, the UV excess may also be contributed by a putative accretion disk that, only in sources with the synchrotron peak shifted at hard X-ray energies, may emerge from the synchrotron continuum.

It is important to note that this method, as applied in this work, contains several simplifications concerning the ambient structures and the corresponding photon fields surrounding the AGN. Specifically, both the BLR and the NLR have been considered as uniform, mono-energetic, and isotropic regions filled with constant photon densities, and their detailed morphology has been ignored. However, this template may be sensibly different in the reality, where for example the density (and the photon density) may vary on parsec scales, or more likely the spectral emission may be spread over a wider energy range than being represented by a mono-energetic radiation. For this reason, this work represents the starting point of a more detailed future study.\\

\subsection{How TeV gamma-ray sources may be affected}
\noindent
The $\gamma\gamma$ absorption process due to the interaction of gamma rays with seed photons of the narrow-line regions may affect notably the observed spectra of TeV gamma-ray sources and their consequent broadband modeling.

\begin{figure}
\centering
\includegraphics[width=0.7\textwidth]{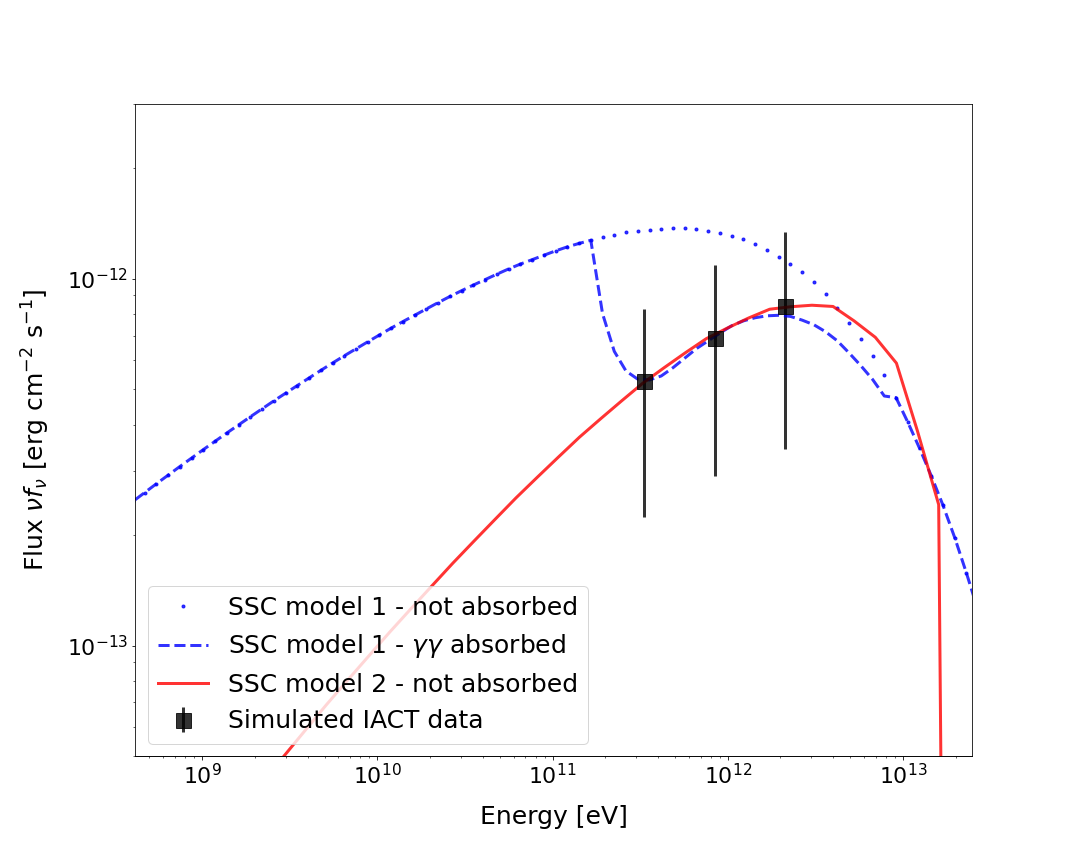}
\caption{Comparison between two different SSC components on simulated IACT data of a hypothetical BL~Lac detected only at some hundreds GeV. The $\gamma\gamma$ absorption modifies sensibly the observed spectrum (already EBL de-absorbed) of model 1. Model 2 represents an incorrect modeling of the source where the absorption has not been considered and the consequent estimation of the peak is shifted at energies $\sim 10$ TeV. More details in the text.}
\label{fig:two_models_SSC_gammagamma_example}
\end{figure}

An example is reported in \Cref{fig:two_models_SSC_gammagamma_example}, where we show the simulated gamma-ray spectrum of a source barely detected by an Imaging Atmospheric Cherenkov Telescope (IACT). In the figure, we show only three data points of the source (already virtually de-absorbed by EBL interaction) between $\sim 300$ GeV and a few TeV. This is a simplification, but it represents a typical case of real observations of TeV gamma-ray sources (as an example, the very same case of PGC~2402248 presented in \Cref{fig:pgc_gammaray_data} if the first and last MAGIC data point were not available), and specifically of observations of EHBLs. These objects are usually very faint and - in order to be detected - they require very long exposure times ($\gtrsim 50$~hours). 

In \Cref{fig:two_models_SSC_gammagamma_example} we show also two different SSC components: SSC~model~2  (red continuous line) is a broadband modeling of the source if we do not suspect any absorption feature affecting the gamma-ray spectrum, and SSC~model~1 (blue dashed line) represents another combination of parameters assuming the presence of an absorption feature. 
SSC~model~1 is shown both with the absorbed component (blue dashed line) and with the intrinsic spectrum before absorption (blue dotted line), which represents the modelled emission. Clearly, the $\gamma\gamma$ interaction modifies sensibly the observed spectral index of the TeV gamma-ray spectrum of the BL~Lac, that is very hard in the blue dashed line rather than flat in the blue dotted line.

On the other hand, if the absorption feature is not suspected, the modeler may apply a \mbox{broadband} model with a much higher peak frequency and a lower maximum luminosity of the bump (compare red continuous line with blue dotted line). This shift of the peak of about one order of magnitude is usually very important on the point of view of the physical parameters needed to produce the model. Specifically in the case of EHBLs, which are characterized by model parameters that are usually very extreme, it may be of fundamental importance in order to reduce the severity of the required physics, and may modify sensibly their sub-classification \citep[see][]{foffano-2019}.

This biased modeling can be avoided only when there is more information on the possible presence of a NLR (in the case of emission lines in the optical spectrum) or when more data points are available at lower and higher energies with respect to the simulated data in \Cref{fig:two_models_SSC_gammagamma_example}. In the case of faint sources (e.g. several EHBLs), only a careful observation strategy and deep exposures of the key instruments may unveil this spectral effect. In fact, at high-energy gamma~rays they are barely detected and usually require long integration time of \emph{Fermi}-LAT (several years). This is unfavourable with respect to the shorter exposures required by IACT observations ($\sim50$ hours), and strictly simultaneous observations may not provide flux data points but only upper limits. On the other side, data points at higher TeV energies are often difficult to obtain because - even if the intrinsic spectrum peaks at extremely-high energies (only a few sources are currently known to be so energetic) - the observed spectrum is severely affected by the EBL absorption and it is usually barely reaching the sensitivity of the current IACTs. In this regard, a key role will be certainly played by the Cherenkov Telescope Array (CTA), which will notably increase the spectral information and the statistics of the most interesting targets.\\

\section{Conclusions}
\label{sec:conclusions}

\noindent
The production site of gamma rays in blazars is closely related to their interaction with the photon fields surrounding the active galactic nucleus. 
In this paper, we studied the spectral effects of $\gamma\gamma$ pair-production process in BL~Lac objects, specifically when highly energetic radiation of the jet interacts with low-energy seed photons emitted by AGN structures close to the jet trajectory. Such process may produce evident features in the gamma-ray spectrum, detectable at a few hundreds GeV. Then, an analysis of these spectral features may indirectly unveil the presence of ambient structures.

An interaction with a putative broad-line region may reduce the gamma-ray flux only if its production site were very close to the central engine.
Alternatively, if the blazar jet interacts with a bath of optical-UV seed photons produced by a narrow-line region extending  a few hundreds parsec from the central engine, this could produce detectable absorption features in the sub-TeV spectrum. However, it is important to notice that this method estimates the total photon column density of the seed photons interacting with the gamma~rays, and then the specific values of size and photon density of the emitting region are not directly constrained. So, this method may describe also non typical structures, and it is applicable to a large variety of AGN environments. 

This absorption effect may be detected only on sources with favourable spectral properties, and that 1) are well detected in VHE gamma rays, 2) show hard spectrum extending up to several hundreds GeV, 3) show a clean spectral shape (without contamination by other spectral features) in that band, and 4) show a relatively stable flux at those energies.

Interestingly, the most promising sources where this effect could be easily detectable are HBLs and EHBLs (\emph{extreme blazars}), that with their emission reaching TeV energies may be powerful probes to investigate the presence of NLR in BL~Lac objects.
In this regard, we discussed the case of the extreme blazar PGC~2402248 (also named 2WHSP~J073326.7+515354) recently detected at TeV energies by the MAGIC Collaboration, which shows an evidence of spectral absorption that may be due to a $\gamma\gamma$ process. In this picture, the seed photons are compatible with those emitted by a narrow-line region located along the jet path (outside the broad-line region), which is illuminated by the jet itself and responsible for the partial absorption of sub-TeV gamma rays.

These spectral features - depending on the specific seed photon fields in each object - may be present in the spectra of a number of BL~Lac objects in general. Also past observations of gamma-ray sources may be interested by such effects. Additionally, since the $\gamma\gamma$ absorption can modify the spectral index at a few hundreds GeV, the broadband modeling of several BL~Lac objects may be severely affected when these absorption effects are not considered.
For this reason, a systematic analysis of the presence of these absorption features in blazars is particularly important, and will be presented in a forthcoming paper.\\

\bigskip
\noindent
{\bf Acknowledgements}\\
We thank the journal referee for the insightful and valuable comments that helped us to improve the manuscript.\\
Research carried out through partial support of the ASI/INAF AGILE contract I/028/12/05. \\
This work made partial use of the software JetSeT \citep{jetset4, jetset3, jetset2,jetset1}.

\bibliography{bib_paper_gammagamma}
\bibliographystyle{aasjournal}



\end{document}